%% file: conselice.tex
\documentstyle[10pt,aaspp4]{article}
\def\deg{$^{\circ}\,$}
\def\solm{M$_{\odot}\,$}
\def\kms{km~s$^{-1}\,$}
\def\hub{h$_{100}$$^{-1}\,$}

\begin{document}

\title{On the Nature of the NGC 1275 System}

\author{Christopher J. Conselice$^1$, John S. Gallagher, III$^2$}

\affil{Department of Astronomy, University of Wisconsin, Madison, 475 N. 
Charter St. Madison, WI, 53706-1582: consel@stsci.edu, jsg@astro.wisc.edu}

\author{Rosemary F.G. Wyse$^2$}

\affil{Department of Physics and Astronomy, Johns Hopkins University,
Baltimore, MD. 21218; wyse@tarkus.pha.jhu.edu}

\altaffiltext{1}{Space Telescope Science Institute, 3700 San Martin
Drive, Baltimore MD, 21218.}

\altaffiltext{2}{Visiting Astronomer, Kitt Peak National Observatory,
National Optical Astronomy Observatories, which is operated by the
Association of Universities for Research in Astronomy, Inc. (AURA) under
cooperative agreement with the National Science Foundation. }


\begin{center}
{\bf {\it Accepted to the Astronomical Journal}}
\end{center}

\begin{abstract}

Sub-arcsecond images, taken in B, R, and H$\alpha$ filters, and area
spectroscopy obtained with the WIYN 3.5-m telescope provide the basis
for an investigation of the unusual structures in the stellar body and
ionized gas in and around the Perseus cluster central galaxy,
NGC~1275.  Our H$\alpha$ filter is tuned to gas at the velocity of
NGC~1275, revealing complex, probably unresolved,
small-scale features in the extended ionized gas, located up to 50\hub
kpc from NGC~1275.  The mean H$\alpha$ surface brightness  
varies little along the 
outer  filaments; this, together with the complex
excitation state demonstrated by spectra, imply that the filaments are
likely to be tubes, or ribbons, of gas.  The morphology,
location and inferred physical parameters of the gas in the filaments
are consistent with a model whereby the filaments form through
compression of the intracluster gas by relativistic plasma emitted
from the active nucleus of NGC~1275.  Imaging spectroscopy with the
Densepak fiber array on WIYN suggests partial rotational support of
the inner component of low velocity ionized gas.  Our broad-band data
is used to derive color maps of the stellar distribution, and also to
investigate asymmetries in the stellar surface brightness.  We confirm
and extend evidence for features in the stellar body of NGC~1275, and
identify outer stellar regions containing very blue, probably very
young, star clusters.  We interpret these as evidence for recent
accretion of a gas-rich system, with subsequent star formation.  Other
star clusters are identified, some of which are possibly associated
with the high velocity 8200 \kms emission line system, being in the
same projected location.  We suggest that two main processes, which
may be causally connected, are responsible for the rich phenomenology
of the NGC~1275 system -- NGC~1275 experienced a recent merger/interaction
with a group of gas-rich galaxies, and recent outflows from its AGN have
compressed the intracluster gas, and perhaps the gas in the infalling
galaxies, to produce a complex web of filaments.

\end{abstract}

\section{Introduction}

NGC~1275 (Perseus A, 3C 84) is one of the most unusual early-type
galaxies in the nearby universe and contains an example of almost every
known extragalactic phenomenon.  However, several of its basic observed
features still remain a mystery.  NGC~1275 is located at the center of
the Perseus cluster and resembles a normal elliptical galaxy on
low-resolution plates (Hubble 1931).   Humason (1932) and later
Seyfert (1943) discovered strong emission lines in NGC~1275.  Later,
Minkowski (1955) found two distinct emission line systems
towards NGC~1275; a high velocity (HV) component at V = 8200 \kms, and a
low velocity (LV) one at V = 5200 \kms.  The stellar radial velocity of NGC
1275 is 5264$\pm$11 \kms (Huchra, Vogeley \& Geller 1999) while the
velocity dispersion of the Perseus cluster is 1277 \kms (Struble \&
Rood 1991). The puzzling nature of NGC~1275 was compounded by the
discovery of an extensive array of emission line filaments projecting
away from the central galaxy (Minkowski 1957; Lynds 1970).  The origin
of these features is still being debated (e.g., McNamara, O'Connell, \& 
Sarazin 1996; Sabra, Shields, \& Filippenko 2000).

The HV system is located to the north and north-west direction
of the nucleus (Burbidge \& Burbidge 1965; Rubin et al. 1977).
The presence of HI and X-ray absorption
clearly places this gas in front of the nucleus, where it is difficult to
explain as a simple outflow (De Young, Roberts \& Saslaw 1973; van
Gorkom \& Ekers 1983; Fabian et al. 2000).  Thus, this object could be 
associated with a 
foreground late-type spiral falling into the cluster
(Rubin et al. 1977; Kent \& Sargent 1979; Boroson 1990; Caulet et al.
1992).

Despite appearing as an
elliptical galaxy on photographic plates, particularly in the red, NGC
1275 has relatively blue central colors (e.g. van den Bergh 1977), massive 
young
star clusters (Shields \& Filippenko 1990; Holtzman et al. 1992), an
A-type integrated blue region spectrum
(Wirth et al. 1983) and internal sub-structures
(Holtzman et al. 1992). Massive, short-lived stars are forming in 
NGC~1275, that perhaps originate from gas accretion induced in a cooling
flow (e.g. Cowie et al. 1980; Sarazin \&
O'Connell 1983), or their formation is induced from a galaxy merger
independent of the high-velocity system (Holtzman et al. 1992).

Is star formation related to the filamentary structure of ionized
gas seen in Figure 1?   Fabian and Nulsen (1977) 
interpreted these filaments as a result of accretion of cooling intracluster 
medium (ICM) gas.   Pioneering
digital optical
spectroscopy yielded emission line intensities of the filaments
that are consistent with
either shocked gas or gas photo-ionization by the 
inferred active BL Lac nucleus in NGC~1275 (Kent
\& Sargent 1979). Recently Sabra, Shields, and Filippenko
(2000) found that neither shocks nor photoionization alone can reproduce the
observed emission line intensity ratios in the inner part of the galaxy.

In this paper, we present high-resolution images, integral field and
long-slit spectroscopy of NGC~1275 taken with the WIYN 3.5m and KPNO 4m
telescopes. We present evidence that the low-velocity ionized gas within
NGC~1275 is partially
supported by rotation, consistent with a rotating gas
disk as seen in emission from CO (Bridges \& Irwin 1998).  Properties of the
filament system as a whole appear to be consistent with the model proposed by 
McNamara, O'Connell, and Sarazin
(1996; hereafter MOS96).  They suggested that the ionized
filaments originate from
interactions between the hot ICM and the
relativistic plasma responsible for the radio source (see also
B\"ohringer et al. 1993; Heinz, Reynolds \& Begelman 1998; Churazov et
al. 2000).   Instabilities
should occur in the compressed and 
cooling ICM as it sinks through the dilute relativistic plasma, 
possibly in a
manner similar to that of ionized gas filaments in the Crab
nebula (Hester et al.  1996). This process could lead to the production of a
radial system of filaments.

We employ several methods to investigate the structure of the stellar
body of NGC~1275, including B$-$R color maps, and the removal of 
symmetric, or smooth light, components to reveal complex or asymmetric
features. We find more
evidence for low amplitude, blue stellar shells or ripples in the main body of the
galaxy (cf., Holtzman et al. 1992; Carlson et al. 1998) and blue star clusters
possibly not associated with the LV H$\alpha$ emission. NGC~1275
clearly suffered a recent, significant perturbation, probably induced
by some type of merger, which may be the source of the system's peculiarities.

Section 2 of the paper describes the observations, Section 3 is an
analysis of the data, Section 4 is an interpretation, while
Section 5 gives a summary. Throughout this paper we use the
following terminology: `low-velocity (LV) system' denotes the 5200 \kms
ionized gas around NGC~1275, `high-velocity (HV) system' denotes the
8200 \kms ionized gas system localized to the northwest of NGC~1275,
and `NGC~1275' refers to the central galaxy (redshift 5264~\kms) of the
Perseus cluster without regard to the LV or HV ionized gas
components. For quantities not expressed in terms of \hub, we adopt a
distance to the Perseus cluster of 70~Mpc.

\section{Observations}

The images used in this paper were taken in 1998 November with the
WIYN 3.5m telescope located at the Kitt Peak National Observatory near
Tucson, AZ.  CCD images were obtained with a 2048$^{2}$-pixel thinned
S2kB device, producing data with a scale of 0.2 arcsec per pixel.  The
field of view of each image is 6.8 $\times$ 6.8 arcmin$^{2}$. NGC~1275
and the surrounding region of the core of the cluster were observed in
the broadband R and B filters, as well as a narrow band H$\alpha$
filter centered on the radial velocity of the LV component (KPNO
filter KP1495 with central wavelength 6690\,\AA\, and FWHM of 77\,\AA)
which excludes most of the emission from the HV system.

Exposure times for the R
images were 2 $\times$ 800 sec, while the exposure times for the B images
were 3 $\times$ 700 sec, and the H$\alpha$ images had exposure times of 3
$\times$ 800 sec.   The average seeing for all the combined images is
0.6\arcsec - 0.8\arcsec. Conditions were photometric, allowing magnitudes 
to be calibrated using Landolt standard star fields. These images are part
of a larger program to study the galaxy populations in the Perseus
Cluster (e.g., Conselice, Gallagher \& Wyse 2001).  A total of 231
arcmin$^{2}$ of area in the core of the Perseus cluster was imaged during this
run. 

We removed the continuum from the combined 2400s H$\alpha$ image 
by subtracting a
R band image normalized with respect to field stars, with an estimated 
accuracy of 7\%.  We calibrated the H$\alpha$
fluxes with the spectrophotometric standard SA 29-130, a DA white dwarf
with spectrophotometry by Oke (1974).  We convolved the spectrum
of SA 29-130 with the filter  transmission function to obtain a
calibrated flux, yielding $2.6 \times
10^{-15}$ erg s$^{-1}$ cm$^{-2}$/(count/s) corrected for Galactic
extinction.  The total H$\alpha$ luminosity of the LV component of NGC~1275 in
the H$\alpha$ + [NII] emission lines is then 4.9 $\times 10^{42}$ erg s$^{-1}$.
No significant H$\alpha$ emission is present in locations beyond
the NGC~1275 filaments.  We correct for Galactic extinction using
E(B$-$V) = 0.17 and A$_{\rm B}$ = 0.71, A$_{\rm R}$ = 0.40 (Burstein \& Heiles
1984).

Integral field spectroscopy of the center of NGC~1275 was obtained
with the Densepak fiber array (Barden \& Wade 1988) on WIYN during
November 1998.  Densepak contains 94 optical fibers arranged in an
area 45\arcsec\, $\times$ 27\arcsec, with each fiber separated by
4\arcsec.  The spectral resolution was 3\,\AA\ ($\sim$140~\kms at
H$\alpha$).  Three 900 second exposures were taken, and later reduced by
the IRAF task DOHYDRA.  A CuAr comparison source was used to produce a
wavelength calibration.  The variable and unpredictable flux
throughput of each fiber makes spectrophotometry with Densepak
unreliable.  Accordingly, we derive only  an
integral field velocity map from the fiber spectra.  We measure and fit the 
low-velocity
H$\alpha$ line in each fiber to obtain a central wavelength and
radial velocity, which we use to produce a velocity field.

Additional long slit spectroscopy was obtained with the RC spectrograph
on the KPNO 4m Mayall telescope on 1999 October 2 for
spectrophotometric purposes.  We chose three different filaments
around the center and 2\arcmin\, from NGC1275.  The KPC-10A grating
was used, giving a wavelength coverage of 4000\,\AA\, with a 2\arcsec\, 
$\times$ 300\arcsec\,
slit, and  7\,\AA\,  resolution.  Two 750s exposures were taken in each of
the three different areas of the ionized filaments of NGC1275.  These
data were reduced with the KPNO long-slit tools, undergoing a trim,
extraction and sky subtraction.  The spectra were then calibrated in
wavelength with an FeAr lamp and flux calibrated by the standard BD+28
4211.  The useful wavelength range of the spectra is 3700\,\AA\, to
7800\,\AA.  We corrected the spectra for interstellar extinction using the 
method of Cardelli, Clayton and Mathis (1989), using E(B$-$V) = 0.17.

\section{Results}

\subsection{H$\alpha$ Structure of the Low-Velocity Ionized Gas}

\subsubsection{The Forms and Origin of the Ionized Gas Filaments}

One of the most remarkable properties of NGC~1275 is the intricate
array of LV H$\alpha$ filaments that can be seen in Figures 1-6.  Our
angular resolution of 0.6-0.8\arcsec\ corresponds to 150-200\hub~pc at
the distance of NGC~1275, and structures exist in these filaments down to this
resolution limit.  Following the description of the NGC~1275 filaments 
by Lynds (1970), extended H$\alpha$ emission
has been observed surrounding other central cluster galaxies (e.g.,
Heckman 1981, Heckman et al. 1989). In some cases filamentary
morphologies similar to that in NGC~1275 are found, as in {\it
Hubble Space Telescope} images of the central cD galaxy in the cluster
Abell~2597 (Koekemoer et al. 1999). 

The filaments cover a total linear distance of nearly 100\hub kpc from
north to south.  These extend
from a central system of filamentary ionized gas in NGC~1275
that is elongated over approximately $\approx$40\hub kpc in the
east-west direction, along the major axis of the stellar part of the
galaxy.  The large north-south extent of 100 \hub kpc occurs for only a
small number of thin, largely unresolved filaments (Figures 2 \& 5).
Figure 7 shows both the surface brightness profile of the H$\alpha$ emission
and its projected filling factor.  Despite the fact that the filling
factor declines significantly after $\sim$ 4 kpc, the annular brightness
of the filaments remains high.

There are two major forms of ionized gas structures around
NGC~1275 - radial and tangential filaments.  The radial types point 
towards the center of NGC~1275; while tangential features usually, but
not always, are located at the outer extreme of the H$\alpha$ emitting
gas in any particular location.  The tangential
filaments curve around the galaxy and are concave towards
its center. How these two types of
filaments relate to each other and to the intergalactic medium around
NGC~1275 provides insights into the origin of the ionized gas.  Figures
3 and 4 show examples of radial and tangential filaments, which can
be seen in abundance throughout the filament system in Figures 1 and
2. The east-west axis of NGC~1275 is less extended in 
H$\alpha$ gas, and shows a predominance of tangential filaments (Figure 4).
Figure 3 illustrates a remarkable filament that remains radial for 30 kpc
before folding backwards.  The smooth, very elongated morphologies
of the filaments
suggest that magnetic fields could play a part in defining their
structures, as they do in the smaller Crab Nebula
ionized filament system (Hester et al.  1996).  In fact, some form 
of additional support, such as magnetic fields, is likely needed to
explain the structures of these filaments.   For a  temperature
of 10$^{4}$ K, the filaments will have an average internal random motion $\sim$
10 \kms, giving a lifetime $\sim$ width/velocity = 0.5 kpc/10 \kms = 50 
Myrs.  Either
the filaments are supported by processes that allow them to survive more than
50 Myrs, or they are continuously being reformed.

Since we probably have not completely resolved most filaments, we are thus
likely underestimating their surface brightnesses.
The evolution of filaments near the center of NGC 1275, which might be
more compressed, could be quite complex. It is possible
that compression and instabilities operate in tandem, as in the Crab
nebula (e.g., Hester et al. 1996). Features such as the long linear
structure of ionized filaments extending to the north could be the
remnants of ICM gas that was displaced by a northward non-thermal radio
jet (see Figure 6) that may have been more intense in the past 
(Pedlar et al.  1990,
MOS96, Blundel, Kassim, \& Perley 2000).  Figure 6 displays the contours from a
1320 MHz VLA radio
map, from Pedlar et al. (1990), over the H$\alpha$ + continuum
image of the central parts of NGC~1275.  This Figure shows the
opposite major axis directions of the two components.
In \S4.2 we discuss the
MOS96 model for filament production in interactions between relativistic
plasma outflows from NGC~1275 and the ICM.

The complex filament morphologies lead us to ask what is the true
distribution of the H$\alpha$ gas around NGC~1275.  Do the filaments
have tubular structures, or are they extended sheets between
ionization and shock fronts, possibly along the edges of expanding
bubbles?  The surface brightnesses of individual filaments are
relatively constant over 50~kpc. For example, the large radial
filament at the base of the `horseshoe' in Figure 3 has 1.0 $<$
I(H$\alpha
+$[N~II])$/$(10$^{-15}$~erg~s$^{-1}$~cm$^{-2}$~arcsec$^{-2}$) $<$ 6,
while a similar range is found along the very extended northern
filaments (Figure 5; Table 1).  The filaments often have high contrast
with respect to their surroundings, again suggesting a relatively
compact physical structure.  Table 1 lists the fluxes and intensities
of the emission line regions labeled in Figure 5, measured using a
1$\arcsec$ circular aperture, relative to a local sky measurement.

The near constant surface brightnesses, large linear extent
of the northern filaments, and small angular sizes of the
radial filaments all imply that they are tubes or compact ribbons of
ionized gas, and not projected edges of large bubbles.  Further support
for the filament interpretation comes from the presence of localized,
unresolved high surface brightness areas.  These ``knots'' could be
examples of filaments seen end on, especially 
in the central part of the filament system 
(see Figure 4 \& 5 and entries 28 - 38 in Table 2 for examples), a similar 
phenomenon is observed in the Crab nebula (see Hester et al.
1996). However, some knots are 
far out in the system where an end-on projection is less likely; 
these could be locations where the volume emissivity is locally
enhanced, e.g., due to high electron densities or more intense
heating of the gas.  These ionized knots, labeled as hexagons in Figure 5, 
mostly have
H$\alpha$ fluxes $\sim$ 10$^{-14}$ erg~s$^{-1}$~cm$^{-2}$.  Some of
these could also be HII regions.

The tangential filaments also present an interesting picture. When
curvature is present, the filaments close in around NGC~1275; an
outstanding example is the `horseshoe' located to the northwest of
NGC~1275 (Figure 3). This pattern is commonly associated with an {\it
outward} displacement of gas; the loops could mark the edges of expanding
regions where gas has been compressed and cooled. The relationship
between the location of the LV system ionized filaments,
disturbed X-ray emitting gas, and the non-thermal radio source all are
consistent with the ionized gas forming where the relativistic plasma
recently displaced the hot ICM (B\"ohringer et al. 1993; Churazov et al. 2000;
see \S 4.1).  It is not clear how these concentric ionized features could be
produced via an inward moving cooling flow, unless they represented
regions where the flow was decelerated by an outward pressure, a model
that is conceptually the same as that for an expanding source.

The widths of the filaments vary between $>$ 0.5\hub ~kpc in clumps of
material, to unresolved features with sizes of  
$< 165$\hub ~pc further out in the cluster.
Any explanation for their origin must account for these small relative
widths of the filaments in comparison to their very long lengths
(ratio width/length $> 100$, especially at 50\hub ~kpc, the largest
distance from the center of NGC~1275 where filaments are seen). 

A simple model based on the measured H$\alpha$ surface brightnesses of
the filaments (corrected for Galactic extinction and assuming that
about half the flux in the filter is from [N~II]) indicate lower bound
electron densities of $\geq$10~cm$^{-3}$. The filaments are then much
denser than the surrounding ICM (n$\sim$10$^{-2}$~cm$^{-3}$) or
relativistic plasma, thus a sinking instability could operate. The
pressure of these filaments is then $\geq$ 10$^5$ K~cm$^{-3}$ for an
electron temperature of $\sim$10$^4$~ K.  This is below the pressure
of the ICM, 10$^6$ K~cm$^{-3}$, thus the filaments could be
experiencing dynamic compression. The inferred presence of a cooling
flow (e.g., Fabian et al. 1981) would show that the ICM is not in
perfect hydrostatic equilibrium, which could lead to pressure
variations over large radial scales, such as the extent of the filaments.  
However, the $[$SII$]$ doublet line ratios from spectra
(Heckman et al. 1989; Sabra et al. 2000), while covering only a few
locations, indicate higher mean electron densities by as much as a
factor of 10, in which case the pressure in the filaments is close to
that of the ICM.  This also would imply a low volume filling-factor
for the ionized gas within the filaments.

\subsubsection{Photoionization}

Shocks or non-thermal radiation from the AGN at the center of NGC~1275
are the most popular explanations for the ionization
source of the filaments (Kent and Sargent 1979, Heckman et al. 1989).
Others have tried to account for the high ionization flux by a merger
between the HV and LV velocity components
(Hu et al. 1983), or processes associated with the ICM, such as X-ray
photoionization or heating by magnetic reconnections (Heckman et al.
1989, Sabra et al.  2000). However, as emphasized by Sabra et al. none of
these mechanisms readily lead to the observed emission line intensity
ratios in the filaments.

Figure 7 presents another way to test ionization models. We plot the
observed surface brightness profile of the H$\alpha$ flux around NGC~1275
and compare it with the predictions of a simple central source
photoionization model, where the rate of ionization equals the rate of
recombinations. We assume for the purposes of this plot a distribution of HI 
that
falls off quadratically with spatial distance from the center; the actual
form we adopt is not important here since, as can be understood from
Figure 7, unless the HI is of uniform density, an unlikely case, this
model fails to provide an explanation for the observed Halpha
distribution.

The observed H-alpha intensity profile, in
elliptical apertures, centered on NGC~1275 is indicated in red in Figure 7. 
The result depends sensitively
on sky subtraction and thus becomes rather noisy at large radii. The
dashed line shows the predicted flux from an isotropic central ionization
source, normalized\footnote{R denotes the projected radius,
while r is the physical radius.} at {\rm R = 0}, without any radiative 
losses other than
r$^{-2}$ dimming. This is an {\it upper} limit to any isotropic central
ionization source.

There is a clear shortfall in the model predictions at large radii. This
is compounded by the fact that the decline in observed mean intensity is
due in large part to a decline in projected filling factor, as shown by
the blue curve and right-hand scale in Figure 7, and not in the decline in 
the mean
intensity of an emitting filament.  From Table 1 we see that
the mean brightness of typical filaments drops by only a factor of a few
between the inner and outer regions of the filament system.  Dust could be
affecting the distribution of H$\alpha$ flux, but not to the extent seen
in Figure 7 since the dust is largely localized near the HV system (see
the B$-$R color map in Figure 8 and Keel \& White 2001).

The H$\alpha$ surface brightness in the filaments is nearly always
significantly higher, sometimes well over a factor of $\sim$100, than that
predicted by a central ionization source model. If the source were beamed,
this problem might be avoided, but but it is not clear how a beamed
photoionization source could power the extensive system of ionized
filaments surrounding NGC~1275.  Furthermore, the presence of bright,
compact emission knots outside of the inner filament system
is difficult to understand in any model with a central source of
photoionization. Hence we conclude, in agreement with Heckman et al.
(1989), that the AGN at the center of NGC~1275 is unable to ionize
the filaments, and it is likely that the ionization source,
while not a pure shock,  is physically associated with the
filaments (see also Sabra et al. 2000).

\subsection{Star Clusters and Stellar Features}

The nature of young star clusters associated with NGC~1275 has been
widely discussed, beginning with the discovery of HII regions by
Shields \& Filippenko (1990) and becoming more vigorous following the
first {\it HST} imaging of this system by Holtzman et al. (1992).
Holtzman et al. (1992) characterized a population of blue, possibly
young, clusters near the core and suggested that some of these have
sizes and stellar masses which would allow them to evolve into
globular star clusters.  Later studies measured
physical parameters of these clusters, such as age and luminosity, to
determine the star and cluster formation rate, and constrain the
physics behind their formation,  e.g., star formation induced from
galaxy mergers, or cooling flows (Richer et al. 1993; Norgaard-Nielsen
et al. 1993; Faber 1993; Brodie et al. 1998; Carlson et al. 1998).

The spectra and colors of the clusters suggest rather short-lived
epochs of cluster formation, rather than the continuous formation
expected in cooling flow models (e.g., Holtzman et al.  1992; Carlson
et al. 1998; Brodie et al. 1998).  Other observations and studies go
further, interpreting the narrow range of colors as implying that
these clusters formed through merger events (Holtzman et al. 1992;
Faber 1993; Carlson et al. 1998).  This conclusion has been disputed
based on other photometric studies (e.g., Richer et al. 1993) that
find a range of colors.  The accuracy of the photometry has however
been debated in the literature (Faber 1993; Brodie et al. 1998).  The
presence of young objects is compellingly demonstrated by the WFPC2
photometry of Carlson et al. (1998), who find a bimodal distribution
of cluster colors with a population of blue clusters with colors
B$-$R $\approx 0.3 \pm$0.2.

Since some uncertainties remain regarding photometry for the brightest
cluster complexes, we have measured B and R magnitudes for the
brighter clusters in the main body and for the asymmetric stellar features in 
NGC~1275.  We also
present the first results for objects in the blue star forming extensions of
NGC~1275. The clusters measured are labeled in Figures 8 and 9, and are 
listed in Table 2. These specific clusters are among the brightest in the 
NGC~1275 system.  These clusters are not located in
regions of strong emission from ionized gas, which would contribute extra
light to both the B and R bands. Line emission is thus unlikely to be a
substantial source of error in our colors, which are dominated by
uncertainties in the complex continuum backgrounds.  We 
therefore do not attempt to correct for any line emission associated
with the individual clusters.

Accurate photometry of the NGC~1275 star clusters is highly compromised
by the `irregular' nature of the highly
structured inner parts of NGC~1275, a result of dust, star formation and 
other features.  To remove
these we first subtract out all asymmetric parts from NGC~1275
and then fit isophotes to the `symmetric' residual image.   We then subtract
out these fitted isophotes from the original image.  Aperture photometry, with
a 1.2\arcsec\, aperture, 
is then performed on the clusters in this subtracted image. 

\subsubsection{Inner Regions}

The inner star clusters whose colors we measure are labeled on Figure 9, 
and listed
in Table 2.  The majority of these clusters have similar colors, with
(B$-$R)$_{0}$ $\sim$ 0.3 - 0.7, with a few blue clusters with (B$-$R)$_{0}$
$\sim$ 0.0.  This is similar to the broad distribution found by Carlson et
al. (1998) for the bright, blue clusters.\footnote{Carlson et al. also
detected a population of faint red clusters that we do not consider
here.}  Photometry of the inner clusters is also presented in Holtzman et
al. (1992), Richer et al. (1993), Norgaard-Nielsen et al. (1993) and
Brodie et al.  (1998). A comparison to these studies can be
effectively done using Table 2 of Brodie et al. (1998).  The rough
corresponding objects in our Table 2, in comparison to those of Brodie
et al. (1998: Table 2), are objects 1, 2, 3, 4, 5, 6 with H2, H3, H4+9, H1,
H6, and H5.  The average absolute differences between our photometry
and the HST photometry of Brodie et al. (1998), using the same
Galactic extinction corrections, is $\sim$ 0.24 mag in the B and R
bands with $\sigma_{B,R}$ = 0.28 and 0.26.  The average color
difference is (B$-$R)$_{\rm Brodie}$ - (B$-$R)$_{\rm This
Paper}$~$\sim$~0.20.  Given the slightly different filter systems and
approaches used to measuring photometry, we a priori would not expect
to measure exactly the same values.  The values are, however, consistent 
within the quoted errors of each measurement. The blue colors
match those expected for rather young, aged $< 500$~Myr, clusters, and the 
implications of a rather homogeneous population of young clusters will be 
discussed in \S 4.2. 

\subsubsection{Outer Stellar Regions}

The pioneering U-band images taken by MOS96 detected extended blue
regions in NGC~1275 that are not associated with the ionized filaments.
We also detect new outer star clusters, shown in Figure 8 and labeled as 
clusters 17 - 22 in Table 2, that are roughly
1\arcmin\, from the center of NGC~1275.
These features clearly extend beyond the main stellar
body of NGC~1275 as seen in Figure 1 and in the B$-$R color map (Figure 8). 
We measure very blue colors with (B$-$R)$_{0}$ $\sim -0.3$ to 0.0 (Table 2),
consistent with an   age of $< 100$ Myr,  younger than  the inner clusters.
However, differential reddening may also be partially responsible for
these color differences, thus we can not rule out the
possibility that they are coeval systems.

The distinction of these clusters from the ionized filaments is demonstrated 
by comparing Figures 2 and
8, where the H$\alpha +$[NII] emission and blue continuum are
shown. The two main regions of extended star formation have approximate
mirror symmetry around the center of NGC~1275, consistent with an
origin from either tidal arms/streamers from a galaxy merger, or effects 
of a bi-polar jet from the nucleus. The loop or `arm' structure in
the southeastern region ``A'' favors the jet model, where stars
could be produced around an expanding cocoon driven by a relativistic
plasma (see \S4.2). The northwestern blue region ``B'' has a linear
structure that could also result from either process.

Region ``A'' in Figure 8
has a projected linear extent of $\approx$8\hub ~kpc and region
``B'' has a projected size of $\approx$15\hub ~kpc. These features are as 
narrow as
0.5~kpc in some places and are often sharply defined. Such crisp structures 
cannot last very long in the absence of additional support, such as
magnetic fields.
For a typical projected distance of 15-20\hub ~kpc for the outer parts of these
features, the orbital period will be $\sim$0.5~Gyr, which sets an upper
limit to their lifetimes.  

The separation of these zones from the current
locations of the bright radio lobes and ionized filaments (MOS96) 
also suggests that if causally connected
the relativistic outflow in NGC~1275 has undergone multiple changes
in direction in the last orbital period, or the filaments formed by some 
other process.
We also note the presence of a redder, fainter, and more diffuse
plume or arrow (Figure 8), extending from the nucleus to the south-southwest. 
This might be an example of a somewhat older jet-induced star forming region, or a remnant of the stellar disk of an accreted galaxy. 

\subsection{The High Velocity System}

The HV (8200 \kms) emission line system can potentially be explained by 
the superposition of a highly
distorted foreground late-type galaxy.  This hypothesis has 
gained acceptance from observational evidence, including the localized 
position of the
HV lines at the northwest corner of NGC~1275 (Rubin  et al.
1977, Caulet et al. 1992), HI and X-ray absorption against the radio source
indicating that the HV system is in the foreground (DeYoung et al.
1973; Fabian et al. 2000), possible rotation of the HV system (Rubin et al.  
1977, Boroson 1990), optical dust lanes and absorption consistent with a 
late-type galaxy (Keel 1983; Keel \& White 2001), an optical spectrum 
consistent with late-type HII
regions without shocks (Kent \& Sargent 1979), and a spectrum with
ordinary stellar Ca II triplet absorption at $\lambda$8498 and Na I D
lines (Boroson 1990) at the same velocity as the HV system.  We also observe
that the star clusters projected towards the HV system have blue 
colors with (B$-$R)$_{0}$ $\sim$ 0.1 - 0.4, possibly indicating an age slightly
younger than 
the clusters associated with the inner LV system clusters 
(Table 2 \& Figure 9).  There are however very blue clusters in the outer
parts of NGC~1275 that are not in the projected location  of the HV system 
(\S 3.2.1; Figure 8).

What is not generally agreed upon is the role this HV
system is playing in the evolution of NGC~1275. 
The distance of the HV material from the center of NGC~1275 is an
important parameter for understanding this problem.  A
significant amount of non-thermal emission is coming from NGC~1275 at the
location of the HV system (Pedlar et al. 1990), and a spatial
correspondence may exist
between the HV and LV emission systems where the HV gas
is present (Unger et al. 1990, Caulet et al. 1992).  H$\alpha$ gas
with velocities between the HV and LV system also exists (Ferruit et al.
1997), suggesting a possible connection.  These observations
allow for the possibility 
that some interaction is occurring between the HV system and the 
main NGC~1275 system (see Boroson 1990 and references therein; Fabian et
al. 2000), 
and led Pedlar et al. (1999) to conclude that
the HV system is probably infalling and within 30 kpc of the NGC~1275
nucleus.

While an infalling galaxy is an attractive model for the HV system, it has
not yet been established how it would fully describe the many peculiarities
of the HV material.  The ionized gas properties of the HV system suggest
that it could be a moderate-size, late-type field galaxy (Unger et al.
1990). However, the estimated rotational velocity of 300~\kms would be high for
such an object.  Others have also argued that the HI gas content is too low 
(van Gorkom \& Ekers
1983) for a late-type galaxy, although gas depletion would be expected for an 
infalling galaxy only $\sim$ 30 kpc from the cluster core. 
The structure of the original HV object is likely further
obscured by perturbations produced by the environment
near NGC~1275.

Other distorted galaxies exist near NGC~1275 (Conselice \& Gallagher
1999), demonstrating that this can be a hostile location for smaller
galaxies. The large inferred cooling flow in the ICM would imply a
higher density ICM
(e.g. Fabian et al. 1981;
White \& Sarazin 1988) offering an efficient mechanism for removing
interstellar gas from an infalling galaxy and thereby further
distorting the system.  For example,
it is possible that a field galaxy interacted
with NGC~1275, leaving behind an infalling trail of stripped gas that
we see as the HV gas system. Such a feature should be short lived;  
whether a tidal tail could survive 
the supersonic passage through the ICM is not obvious, thus models where 
the gas is strongly bound, such as in a galaxy, perhaps are better 
candidates for explaining the HV system in a collisional model.

Recently, Dupke \& Bregman (2001) discovered 
X-ray emission to the east and south of NGC~1275 redshifted by 
several thousand km~s$^{-1}$ with respect to the cluster's central velocity. 
This material therefore overlaps in velocity with the
HV system, but perhaps not spatially.  Dupke \& Bregman interpret the large
peculiar velocities 
as evidence for an off-center merger in the Perseus cluster by a 
group or subcluster. However, they do not consider the possibility 
of a connection to the HV system, whose characteristics also suggest an
association with some type of merger process
that could be a source of the HV material.

\subsection{Motion of the Low Velocity Ionized Gas}

The Densepak integral field unit on the WIYN telescope was used to map
the LV gas velocity field within NGC~1275 (see \S2).  Figure 10 shows
the outlines of the positions of the fibers on Densepak over the
central 30\arcsec $\times$ 40\arcsec\, of NGC~1275, while Figure 11
shows the measured velocity distribution.  Table 3 lists the positions
and velocity values of the gas observed in each fiber.

The velocities across the central 0.33\arcmin\, of the ionized gas
around NGC~1275 range from 4800~\kms to 5500~\kms.  A rough velocity
gradient can be seen across the area, with higher velocities in the
northeast, about 20\arcsec\, north of the center of NGC~1275, while
there is a minimum velocity near 5000 \kms in the southwest area\footnote{This
gradient is difficult to see in gray scale versions of this figure.}.
The average velocity in each indicated 10\arcsec$^{2}$ region is also
indicated.  
The ionized gas in the inner 30\arcsec\, of NGC~1275 could have either a 
component
of rotation about an axis with a position angle of $\sim$ 120\deg, or
is involved in a bipolar outflow/inflow.  This is similar to the position
angle of one of the inner streamers and roughly consistent with
rotational flattening responsible for the east-west orientation
of the ionized gas complex near the center of NGC~1275.  This elongated
structure has earlier been recognized as a possible consequence of
rotation (e.g. Cowie et al. 1983). Bridges \& Irwin (1998) showed
that NGC~1275 contains about 10$^{10}$ \solm of
molecular gas in its inner parts, with a
similar velocity field and spatial distribution
to those of the much smaller mass of ionized gas studied here.

The alignment of both the ionized and molecular inner gas along the
major axis of NGC~1275, roughly orthogonal to the direction of the
radio jets, the large mass of gas in this region, and its velocity
amplitude suggest that this gas is a rotating structure rather than an
outflow.  The rotation amplitude of 250 \kms over a radius of
$\sim$7~kpc implies a dynamical mass of 10$^{11}$ \solm.  The
angular momentum of the inner ionized gas appears to be 
comparable to that of a typical giant spiral galaxy. 
While this might result
from a cooling flow (e.g. Fabian \& Nulsen 1977; Mathews \& Bregman
1978) or from a merger with a galaxy, it is not obvious 
in either case whether or not the gas could align itself along the 
major axis of NGC~1275 sufficiently rapidly.  Whatever the
origin of the central gas complex in NGC~1275,
it clearly has existed with little or no radial
motion for a sufficient time to allow it to orbit NGC~1275 and form a partially
relaxed system near the center of the galaxy; the orbital time of
$\sim$100~Myr sets the minimum time scale for this event.  Like previous
authors, we find a complicated velocity field, suggesting that the
inner gas in NGC~1275 is perturbed, possibly by the AGN (Heckman et al.
1989; Bridges \& Irwin 1998) or by recent merger(s).

\section{Discussion}

\subsection{Excitation of the Low-Velocity Emission Lines}

Emission line intensities from the LV system present as much of a problem
as does its spectacular morphology. Optical spectroscopy of NGC~1275 by Kent \&
Sargent (1979) led them to conclude that the LV filaments could be powered
by either photo-ionization from the active BL Lac nucleus, or result from
shocked gas. Further analyses by Heckman et al. (1989) and Sabra et al.
(2000) also suggest that a combination of shocks and photo-ionization
could contribute to making the NGC~1275 filaments, but that neither of
these processes alone would suffice.

A key feature is the wide range in filament ionization levels, with
pronounced emission from $[$OI$]$ and $[$NI$]$ coexisting with $[$OIII$]$
and $[$NeIII$]$.  Figure 12 shows an example of our Kitt Peak spectra of
the NGC~1275 filaments. Since the ionization levels of H and O are linked
by charge exchange reactions, the presence of strong [OI] emission proves
that warm neutral gas is present in the filaments. By analogy with
conditions in the Crab nebula (Sankrit et al. 1998), we are likely seeing
ionized sheaths surrounding neutral filament cores.  Thus the density
structures of the filaments could play an important role in determining
the emitted spectrum, and may be obscuring the ionization mechanisms.

Further complications arise from the short recombination and cooling time
scales for the filaments. The recombination time is 
$\sim 10^5$ yr/(n$_{\rm e}$ cm$^{-3}$) and the cooling time of the 
ionized gas, if anything,
is shorter than this (cf. Osterbrock 1989), so there must be a power source to
keep the filaments ionized, or they must be re-formed on comparable
time scales.  However, the cooling time for the hot ICM near NGC~1275 is
$\approx1$~Gyr (e.g., Fabian et al. 2000; Dupke \& Bregman 2001). In the
(unlikely) circumstance of no additional heating, the filaments would
become neutral much more rapidly than they could be replenished from the
ICM. The presence of a spatially coherent system whose extent in light
years is larger than their $\sim$10$^4$~yr cooling time would be a major
puzzle.  Evidently the evolution of the filaments is slowed by a source of
heat.

How was gas compressed by factors of $>$10$^3$, to the
point where it becomes neutral, in a location where external heating is
obviously present?  One example of such a process operates in the Crab
nebula to create the well-known network of thermal gas filaments (Sankrit et
al. 1998) thus the situation is not an impossible one.  The ICM around
NGC~1275 is the most likely source of gas for the filaments, but then what
fills the role of the expanding relativistic plasma in the Crab to cause
the ICM in the Perseus cluster to condense into filaments?

Our working hypothesis for filament formation follows the suggestion of
MOS96 that the gas filaments around NGC~1275 are produced by an
interaction between the ICM and relativistic plasma ejected from the
active nucleus of NGC~1275.  In this ``Crab-like'' model, the relativistic
plasma has (or had)  a higher pressure than the ICM and therefore
displaces it. How this occurs depends on the expansion velocity of the
relativistic plasma; if it is supersonic relative to the ICM, then we
would expect shocks to form at the ICM/plasma interface, but the ICM would
be compressed even if the expansion were sub-sonic.

\subsubsection{Relativistic Plasma-ICM Interactions}

Interactions between jets of relativistic plasma and ambient gas in and
around radio galaxies are relatively common; e.g., in Minkowski's object
(van Breugel et al.  1985), Hydra A (McNamara et al. 2000) and Centaurus A
(Graham \& Price 1981; Morganti et al. 1991; Graham 1998; Mould et al.
2000), where they produce star formation and optical emission line
regions.  Relativistic plasma ejected from the nuclei of radio galaxies
also affects the X-ray emitting thermal gas in several galaxies (e.g.
B\"ohringer et al.  1993; Carilli et. al. 1994; Rizza et al. 2000;
McNamara et. al. 2000).

X-ray and radio maps of NGC~1275 (B\"ohringer 1993, Pedlar et al. 1990,
MOS96, Fabian et al. 2000) indicate this type of event may have occurred
around NGC~1275.  These maps reveal that the radio plasma seems to have
displaced the X-ray emitting gas around the northern and southern bases of
the filament system. The linear extension of filaments to the north aligns
with faint emission seen at low frequencies, possibly the remnants of a
former jet structure (Blundell, Kassim, \& Perley 2000).  The H$\alpha$
emission largely avoids both the current radio lobes and the surrounding
X-ray bright areas, but often are exterior to these regions, as expected
if they form in displaced gas.

A simple model for this type of interaction is that a near spherical
bubble of relativistic gas expands from the galaxy into its surroundings.
This bubble or ``cocoon'' is powered by a jet from the active galactic
nucleus (Pedlar et al. 1990; Heinz et al. 1998).  The kinetic energy input
to the relativistic plasma is estimated to currently be about 10$^{43}$
erg s$^{-1}$ in NGC~1275 (Pedlar et al. 1990), similar to the energy
released from the cooling ICM gas. It is therefore energetically possible
for the relativistic plasma to influence the ICM.  The size of the cocoon
at the time it detaches from the jet can be estimated, following Churazov
et al. (2000), to be about 50~kpc in NGC~1275.

Possibly a cocoon displaced the hot ICM surrounding the center of NGC~1275
(B\"ohringer 1993, MOS96).  In principle an expanding cocoon could
produce large scale shocks, a key aspect of the Heinz et al. (1998) model.
Recent observations with {\it Chandra} of NGC~1275 reveal neither
evidence for strong shocks between the radio lobes and the ICM, nor
indications that the radio lobes are expanding supersonically (Fabian
et al. 2000).  From these
data and the results of Sabra et al. (2000) it seems unlikely that shocks
are currently the major factor in the filament system.  However, shocks
could have occurred in the past if we are witnessing a late phase of the
interaction between the radio lobes and the ICM, after the main expansion
stage, as in the evolutionary models of NGC~1275 presented by Reynolds,
Heinz, \& Begelman (2001).

Clearly the production of filaments involves complex physical processes,
and more so if, as is likely, magnetic fields are a significant factor.
However, once cool material is produced, it will be much denser than its
surroundings. This material is likely to be subject to Rayleigh-Taylor and
Kelvin-Helmholtz instabilities, which will act to further concentrate the
cooling gas (e.g., B\"ohringer \& Morfill 1998). The combination of
large scale gas compression from supersonically expanding radio lobes and
subsequent instabilities can plausibly lead to the spectacular NGC~1275
filament system as suggested by MOS96.

While the conceptual model of a radio lobe-ICM interaction producing
filaments is attractive, some important issues are unresolved. Most
critical is whether the hot ICM can be sufficiently compressed along the
boundary of a cocoon so as to cool rapidly enough to create the ionized
filaments.  For an intrinsic ICM cooling time near the main body
of NGC~1275 of $<$1~Gyr (Fabian et al. 2000), we require compression
factors of $\sim$10 for the gas cooling time to be shorter than the
dynamical time scales of $\approx$100~Myr (see \S 3.1.1). A strong adiabatic 
shock
produces a factor of 4 increase in density. Thus, with even modest
post-shock cooling, the compressed gas will be able to drop to temperatures
of $\leq 10^6$~K during the required time interval. At this point the
cooling rate greatly increases due atomic line emission, and the thermal
time scales decrease by at least a factor of 10.  In addition, the expanding
relativistic gas could uplift cool, dense material from the inner
parts of the system, as suggested by Churazov et al. (2000), and thereby
provide another source of dense material for making filaments.

In this model the filaments come from the ICM, thus it can be tested by
comparing the chemical abundances of species in the filaments and the ICM.
If this model were accurate the two abundance measurements would be similar. 
Based on the Perseus cluster ICM chemical
abundances measured by Dupke \& Arnaud (2001), we would expect the
filaments to contain gas with about half of the solar oxygen abundance.
Another test can be made by estimating pressures within the filaments. If
they form from instabilities in the ICM and are long-lived, then they will
be in approximate pressure equilibrium with their surroundings. As we saw
in \S3.1.1, if the electron densities are as high as the allowed upper
limits of $n_e \approx$100~cm$^{-3}$ given by the $[$SII$]$ doublet
emission line ratios, then the pressures could be roughly equal between
the filaments and ICM.  We conclude that current observations neither
eliminate this model, nor suffice to fully establish its veracity.

\subsection{Recent Galaxy Accretion in the NGC~1275 System}

\subsubsection{The LV System}

While NGC~1275 has neither the colors
nor the spectrum of a classical elliptical galaxy (e.g. van den Bergh 1977;
Wirth et al. 1983; Romanishin 1987),  it does have an
R$^{1/4}$ radial surface brightness profile, typical of giant
ellipticals, in its inner regions (e.g., Schombert 1986; Prestwich et al. 
1997).  However, an r$^{1/4}$ profile can form on short time scales
during mergers between disk galaxies, and is often
found in recent mergers, such as Arp 220 and NGC 7252 (Schweizer 1998).
Therefore this feature alone does not tell us if NGC~1275 is fundamentally an
ancient elliptical galaxy.

Holtzman et al. (1992) presented convincing evidence, from images taken with 
the Wide Field Planetary Camera 2 (WFPC2) on the {\it
Hubble Space Telescope}, that NGC~1275
contains complicated internal stellar structures.  They found
what appears to be a spiral arm, or a diffuse asymmetric light
feature, near the center of NGC~1275, as well as
two ripple patterns to the southeast of the center of the galaxy.  
We detect in our data two new ripple features (or rings), for a total
of four (see Figure 8).  We further identify two diffuse ``arms'' 
(Figure 8 \& 9), one more than Holtzman et al. (1992), 
with an additional third small ``arm'' in between.  These ``arm'' features 
have a surface brightness $\mu_{B}$ $\sim$ 22 mag arcsec$^{-2}$,
with moderately blue colors, (B$-$R)$_{0}$ $\sim$ 0.30, suggesting that the
stars in the `arms' are the same age as the clusters.  The magnitude
of each arm is about B = 20.6, or M$_{B} \sim -14$.  These `arms' could have
formed from gas imported from the accreted galaxy, while the
ripple features are potentially from the old stellar component.  The 
luminosities of the arms suggests that the accreted object could be
either a dwarf galaxy, or disrupted young luminous star clusters.

The small scale features in the main body of NGC~1275 support the idea
that this galaxy experienced a merger with a galaxy that contained a
non-negligible stellar mass, but likely with a significantly lower
mass than NGC~1275 (cf. Schweizer \& Seitzer 1988).

Carlson et al. (1989) describe a large population of blue, B$-$R
$\approx$ 0.3, star clusters with ages $\sim$300 Myr.  There is very
little spread in the colors of these bright star clusters, hence they
could have all been created at nearly the same time, although
uncertainties in internal extinction and background subtraction make
precise age-dating of the star clusters difficult (Holtzman et
al. 1992; Carlson et al.  1989).  Corrected for Galactic extinction,
but uncorrected for internal extinction or background light, we find
that most star clusters in this region have (B$-$R)$_{0}$ $\sim$ 0.3 - 0.7,
and absolute magnitudes M$_{\rm B}$ $\sim$ $-12$ to $-15$, corresponding to
ages of $<$1~Gyr for metallicities of $>$0.1 solar (Kurth et
al. 1999).  Based on these ages, it seems likely that the most
recently completed merger in NGC~1275 took place about 300~Myr ago.

We find further support for this merger being either a recent minor or
old major one, in a relative sense, by the use of a color-asymmetry
diagram (CAD).  The CAD is a diagnostic tool for determining physical
properties of galaxies based on the quantitative parameters of
asymmetry and color (see Conselice 1997 and Conselice et al. 2000a).
Briefly, this approach takes advantage of the natural correlation
between mean color and degree of organization in normal galaxies that
is broken when a galaxy is disturbed by interactions, which produce
higher levels of disorganization than in a normal galaxy of the same
color. The organization of the galaxy is measured in terms of an
asymmetry parameter (Conselice et al. 2000a).

Among other properties, many nearby galaxies that do not fit on or
near the CAD normal galaxy fiducial sequence (dashed line in Figure
13) are undergoing an interaction or recent merger (Conselice et al.
2000a,b).  NGC~1275, within a radius that excludes the HV system, has
an asymmetry value of $A = 0.16 \pm 0.01$.  Asymmetries are computed
by rotating an image by 180\deg, subtracting the rotated image from
the original, and then by taking the ratio of the absolute value of
the sum of these residuals to the total flux of the galaxy. NGC 1275
has a color $(B - V)_0 =$ 0.72$\pm$0.03, corrected for internal and
galactic extinction (de Vaucouleurs et al. 1991).  These asymmetry and
color values do not put NGC~1275 in the area of major mergers (Figure
13, Conselice et al.  2000b), but its asymmetry is somewhat large for
what is clearly an underlying spheroidal galaxy, indicating the
presence of the faint peculiar rings representing a small perturbation
in the mainly symmetric galaxy.  The other galaxies in Figure 13 have 
asymmetries measured at larger radii, where mixing
times are longer, than where we measure NGC~1275's value.
NGC~1275 is thus even more asymmetric, in a relative
sense, than that implied by examining its position on Figure 13.

The broad range of estimated clusters ages around NGC~1275 (Richer et
al.  1993; Kaisler et al. 1996) possibly requires an extended star
formation history that might not be associated with recent mergers.
In the future wide-field high resolution images and integral field
spectroscopy of NGC~1275 will allow closer examination of this
formation process through a determination of its star cluster
formation history.

\subsubsection{The HV System}

In Figure 8, we note the area of the HV foreground `galaxy' in our
color image of NGC~1275.  This appears as extra blue light in the
northwest quadrant of NGC~1275, where MOS96 also noted a possible
coincidence of blue light with the HV emission line system.  While we
cannot rule out the possibility that some of the excess light in the
northwest quadrant is from star formation in the LV system, its
location suggests a possible connection with the HV system.

This asymmetric components in this area appears bluer than the rest of
NGC~1275 (MOS96) as is readily seen in the B$-$R color map.  If this
is the possible HV foreground galaxy, then it contains several thick
dust lanes (Keel 1983; Keel \& White 2001) and star clusters. These
clusters are not present in the H$\alpha$ image of the LV emission
from NGC~1275, although some of these clusters are at the LV system
redshift of $\sim$ 5200 \kms (Carlson et al. 1998; Brodie et
al. 1998).  Confirmation of which star clusters are in the HV system
will require further spectroscopy for radial velocity measurements.

The non-detection in the LV H$\alpha$ and the spatial coincidence with
the HV system lends credibility to the possibility that some of the
star clusters and the underlying starlight could be the remains of the
galaxy responsible for the HV emission.  The blue light that could be
part of the HV system does not resemble any normal galaxy.  The
star-forming knots are aligned parallel to dust lane features (Figure
8), suggestive of the pattern seen in the arms of spiral galaxies,
while the distorted shape of the overall system suggests that
dynamical disruption has occurred.  Keel and White (2001) further
suggest that the highly correlated spatial positions of the dust lanes
near the HV system and clusters are strong evidence for an
association.  In any case, whether LV or HV, these star clusters have
essentially the same colors as the spectroscopically confirmed (Brodie
et al. 1998) LV clusters (Table 2).  Interpreted as similar ages, this
would imply a coincidence in the epoch of cluster formation.  This
could arise as a result of an interaction between the gaseous
components of the LV and HV systems, perhaps indicated by the presence
of H$\alpha$ gas at velocities bridging those of the two systems
(Ferruit et al. 1997).

Cluster formation in a burst of star formation associated with the
infall process is also plausible.  It would be useful to seek older,
redder clusters with high radial velocities as a means to identify an
underlying old HV system stellar population that would be an
unambiguous signature of a galactic source for the HV stars, versus
for example, star formation stimulated by a jet.

The interaction of a late-type field spiral falling towards NGC~1275
would be further complicated by the effects of gas stripping
mechanisms. In addition to the usual ram pressure and other gas
removal mechanisms associated with a rapid passage through the ICM
(whose density would be boosted by the cooling flow), an infalling
galaxy also could interact with the expanding relativistic plasma from
the AGN in NGC~1275 (Churazov et al.  2000).  These stripping
processes also offer an explanation for the depleted level of HI gas
in the HV system found by van Gorkom and Ekers (1983). If some of the
gas forms stars during the interaction/stripping, which must be
associated with shocks given the clearly supersonic relative
velocities, then the HV stellar populations produced would have
unusual spatial distributions and should be no older than the time
since the infalling galaxy entered the cluster.

In summary, our observations allow a model where the HV system towards
NGC~1275 was produced by an infalling galaxy.  Identification of
pre-infall HV stellar populations or star clusters with ages of $>$few
Gyr would be consistent with this hypothesis. Until then the
possibility remains that the HV system formed by a more exotic
process, such as an interaction between a jet from the AGN and
something else (galaxy, gas cloud) in the vicinity of
NGC~1275. However, an additional hint that the HV system might be part of a
single event comes from the high redshift X-ray emission component in
the Perseus cluster found by Dupke \& Bregman (2001).

\section{Summary and Discussion}

In this paper we present new observations of the bizarre NGC~1275
system in the Perseus cluster of galaxies.  High quality WIYN
Telescope images of the LV system of ionized gas filaments allow
important structural distinctions to be made.  We show that filaments
have roughly constant surface brightness along their lengths,
inconsistent with photoionization directly from the AGN (see also
Heckman et al.  1989; Sabra et al. 2000).  Tangential filaments
generally curve around the central galaxy, suggesting gas compression
by an outflow from NGC~1275 rather than cooling in a pure inflow.
Filaments also show rich structures, including bright knots, to the
limits of our angular resolution ($\sim$150\hub ~pc).

We follow the suggestion of MOS96 in interpreting the tangential
filaments as regions where the expanding relativistic plasma from the
AGN in NGC~1275 displaced and compressed the ICM, which then rapidly
cooled (B\"ohringer et al.  1993; Heinz et al. 1998; Churazov et al.
2000; Fabian et al. 2000).  Gas along the boundaries of the radio
lobes may be further concentrated by Kelvin-Helmholtz instabilities
and radial filaments can form from the actions of additional cooling
and Rayleigh-Taylor instabilities.  The resulting filaments are
expected to consist of tubes or ribbons, likely with complex
substructure and small ionized gas filling factors.  If this picture
is correct, then we expect the gas in the filaments to have the same
abundances as the ICM from which they formed, and to be in approximate
pressure equilibrium with their surroundings.  Abundances, excitation
and ionization states for tangential and radial filaments should also
be similar to one another at a given distance from the center of
NGC~1275. Problems remain with this model, for example it does not suggest
how long the filaments should last.  The lifetime would depend on
 how stable the radio output 
is, where the ionization comes from, magnetic fields, etc., all things
we do not completely  understand yet.
Future measurements of the internal velocities of
individual components of the filaments will provide useful
insights into these problems. For near pressure equilibrium models, internal 
speeds will
be roughly sonic, while bulk motions can be much faster, albeit still
sub-sonic with respect to the medium in which the moving filaments are
embedded.

The source of filament ionization is not completely understood, but is
unlikely to be a simple result of either shocks or photoionization
(cf., Sabra et al.  2000). However, any heating process cannot be too
effective since the [OI] and [NI] emission indicate the presence of
neutral gas within the filaments.  It is therefore possible that the
filaments hide a significant gas mass within their neutral
cores. However, we find only a few examples of blue stars associated
with ionized filaments, thus they generally are not so dense as to
support extensive star formation.

Our data agree with the MOS96 finding that in the past, dense
filaments probably were sites of star formation outside of the main
body of NGC~1275.  Two regions containing young blue star clusters
extend over $\sim 5$~kpc on either side of the main body of NGC~1275.
The existence of star formation in gas compressed by the relativistic
plasma of a radio source would not be unique to NGC~1275, but we do
not yet properly understand this process. It is not obvious why
extensive star formation occurred in some regions in the relatively
recent past ($<$1~Gyr on the basis of B$-$R colors and dynamical
evolution time scale arguments), but not at the present time.

We further demonstrate that morphological properties of the inner
portions of NGC~1275's ripples (or rings; Figure 8) are consistent
with a recent minor merger (see Schweizer \& Seitzer 1988). From the
sizes of these tidal features we estimate that the merger is no older
than a few $\times 10^8$ years, the same age as the blue star
clusters, and thus a likely cause of the star formation, blue colors
and A-type spectrum seen in the central regions of NGC~1275.  Rich
star clusters are frequently observed in interacting/merging galaxies
(e.g., Whitmore 2000). Moreover,  gravitational instability can occur within 
tidal tails (e.g., Duc and Mirabel 1999) perhaps providing an explanation for 
the spatially extended blue stellar features we identify. 
If this past event also produced an outburst
from the AGN, it could unify the properties of NGC~1275 by creating
radio lobes, and therefore ionized filaments.

Using the WIYN integral field spectrograph, we measure the velocity
structure of the inner part of the LV ionized gas, finding evidence
for rotation in an otherwise chaotic velocity field, where the
velocity spread is $\approx$250 \kms. The velocity field in the inner
ionized gas is similar to that of the much larger mass of molecular
gas in the same area (Bridges \& Irwin 1998). The east-west extension
of the inner complex of ionized gas could reflect the influence of
rotational flattening, suggesting that this gas reservoir is partially
relaxed.  It is not clear where the huge amount of cool ISM within
NGC~1275 came from. If it is due to the cooling flow, then a process
must exist to align the gas angular momentum vector with the minor
axis of the stellar body of NGC~1275 before the gas has settled into a
dynamically cool disk. The large amount of dust in this material also
requires that extensive star formation took place to pollute what
should have been relatively dust-free gas from the ICM. If a
significant part of the ISM in NGC~1275 was supplied by mergers, then
the merged galaxies must have been extremely gas rich, or were
possibly a small group of gas rich systems, that could also produce
the large velocity gradient seen in the Perseus cluster X-ray gas by
Dupke \& Bregman (2001).

  The excess stellar surface brightness
distribution near the HV ionized is distorted and asymmetric with blue
optical colors, as previously discussed by MOS96 and others.  The HV
system could therefore result from a galaxy interacting with NGC~1275.
Star formation in the stripped ISM could be triggered by this
interaction, further complicating the structure of this
region. Additional radial velocity measurements of the excess blue
starlight and a full range of star clusters can provide sharper tests
of the infalling galaxy model.  If the HV system is associated with
the remains of an infalling galaxy that is not yet completely
disrupted, then NGC~1275 recently has accreted at least two galaxies
-- one for the HV system interacting with the ICM of the cluster, and
at least one more to make the ripples, rings and star clusters in its
main body. This would further suggests that we are seeing the final
demise of a small group of galaxies that entered the Perseus cluster.
Other galaxies with positions and velocities close to the HV system
would be further evidence of this. Hierarchical clustering models of
structure formation lead to the formation of clusters of galaxies by
the merging of galaxies, so some ongoing accretion/assimilation may 
be expected. 

While tentative explanations exist for several aspects of NGC~1275,
others remain as mysteries. It is not clear why the core of the
Perseus cluster should host strong interactions at the present epoch,
especially since the remainder of the cluster is remarkably free from low 
velocity galaxy interactions and starbursting galaxies
(Conselice, Gallagher, \& Wyse 2001).  Another problem is the presence
of apparently linear dust lanes (Figures 1, 2 \& 8), including large
clumps of dust near the HV system not seen in other areas.  If the HV
system's gas and stars are being distorted by an interaction with the
ICM and expanding energetic radio plasma, then how does the dust
survive?  Furthermore, neither the role, nor the extent of the
cooling flow in Perseus is fully understood. This leads to unresolved
questions as to how the cooling flow affects the various components of
the cluster, particularly its inner parts in the critical region
around NGC~1275 (e.g., Fabian et al. 2000).

Further observations of the Perseus cluster with X-ray satellites such
as {\it Chandra} and XMM to obtain higher resolution imaging and
detailed abundance measurements of the ICM from X-ray spectroscopy
should clarify this and several other issues. Additional studies at
other wavelengths are needed, including efforts to better define the
characteristics of the young clusters, especially their ages and
kinematics, as well as high angular resolution observations designed
to explore the internal structures of the H$\alpha$ filaments.
Gaining a proper description of what is happening in and around
NGC~1275 remains an important milestone for understanding the
evolution of central galaxies in rich clusters.

We thank the ``WIYN Crew'' for providing a superb and smoothly
operating observatory, and Birgit Otte for assistance on several
technical aspects of this research.  Annette Ferguson kindly provided
the opportunity for us to take long-slit spectra during a run at the
KPNO 4m.  We also thank Lisa Frattare for creating the color image of
the ionized gas around NGC~1275 with our WIYN images (Figure 1), and
John Hoessel and an anonymous referee for useful comments.  This work was 
supported in part by
the National Science Foundation through grants AST-9803018 to the
University of Wisconsin and AST-9804706 to the Johns Hopkins
University (RFGW). JSG appreciates support provided by the Vilas
Trustees through the University of Wisconsin Graduate School. CJC also
gratefully acknowledges support from the Wisconsin/NASA Space Grant
Consortium, a Grant-in-aid of research award from the Benjamin
A. Gould fund from Sigma-Xi and the National Academy of Sciences and a
Graduate Student Research Program (GSRP) Fellowship from NASA.

\begin{figure}
\plotfiddle{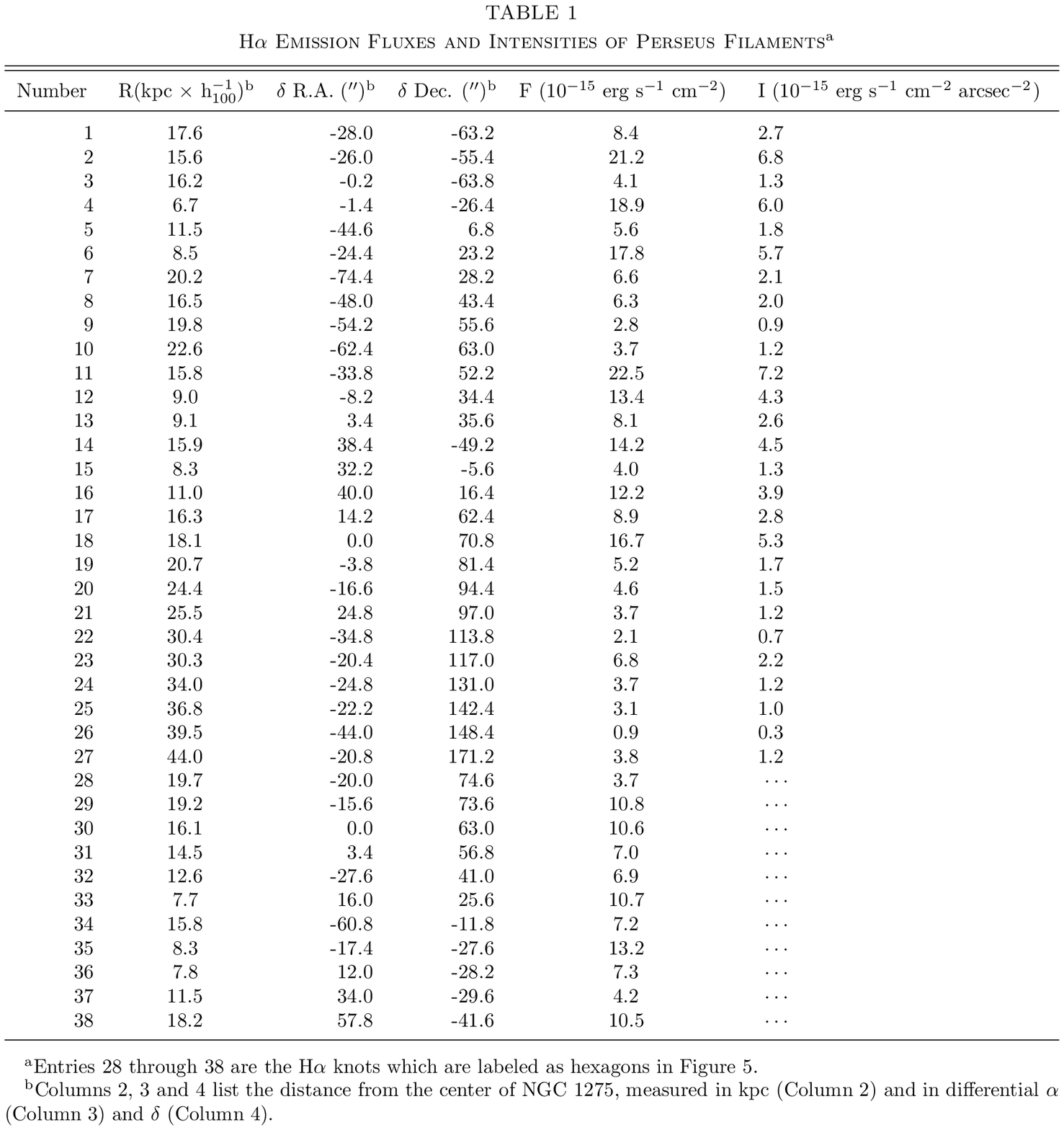}{6.0in}{0}{100}{100}{-310}{-170}
\end{figure}

\begin{figure}
\plotfiddle{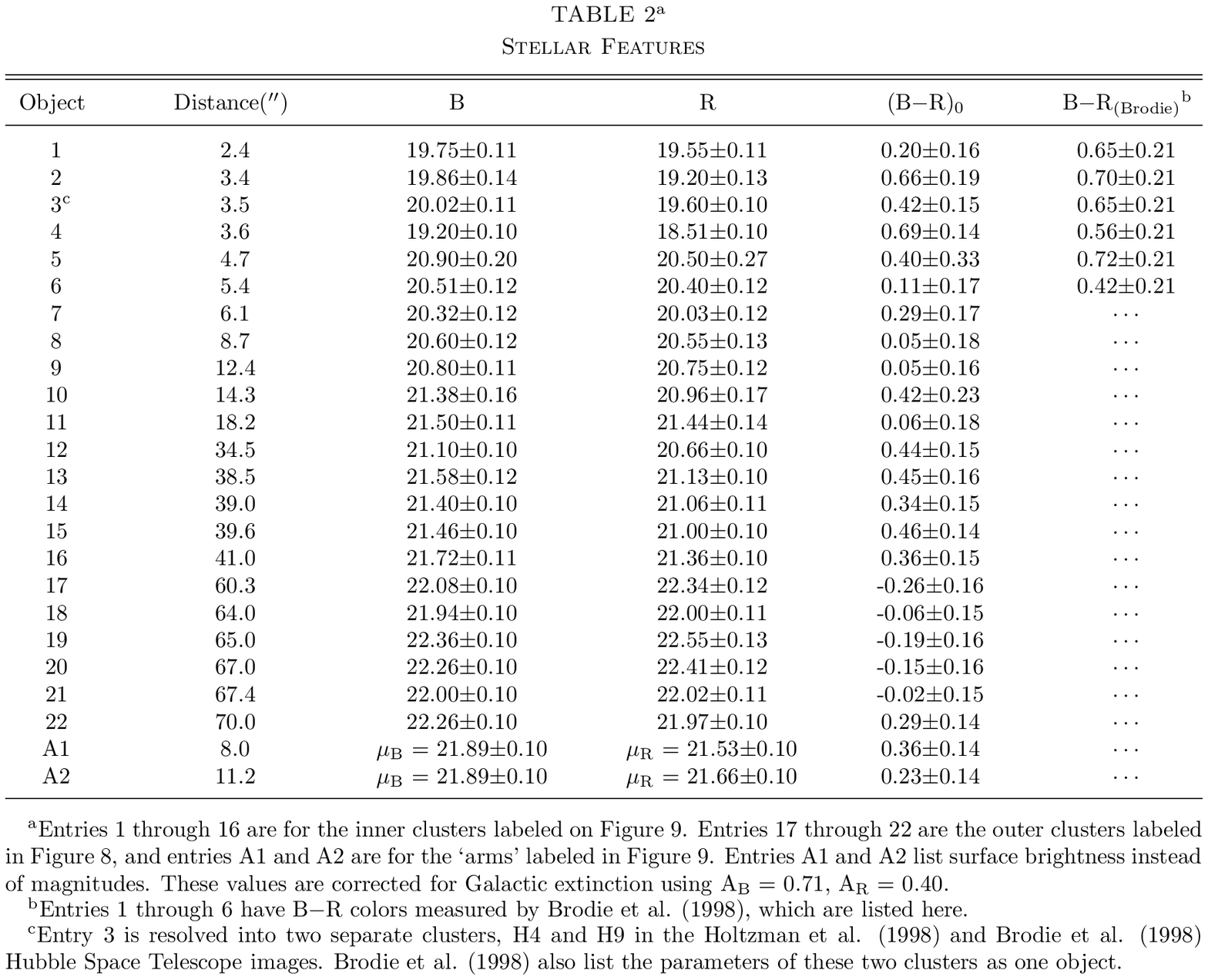}{6.0in}{0}{100}{100}{-310}{-170}
\end{figure}

\begin{figure}
\plotfiddle{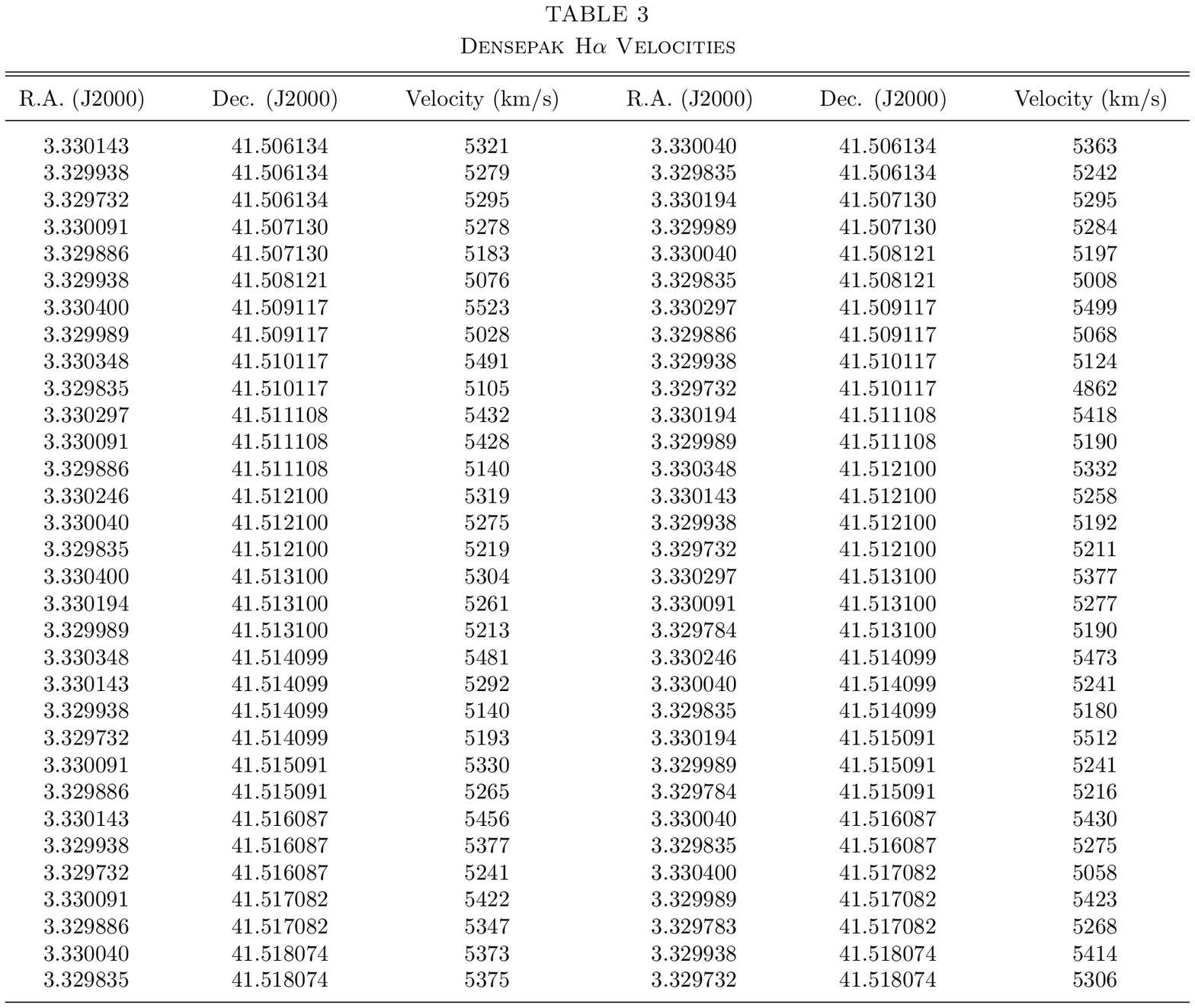}{6.0in}{0}{100}{100}{-310}{-170}
\end{figure}

\clearpage
\input figures_input.tex

\end{document}

%% file: figures_input.tex
\setcounter{figure}{0}

\begin{figure}
\plotfiddle{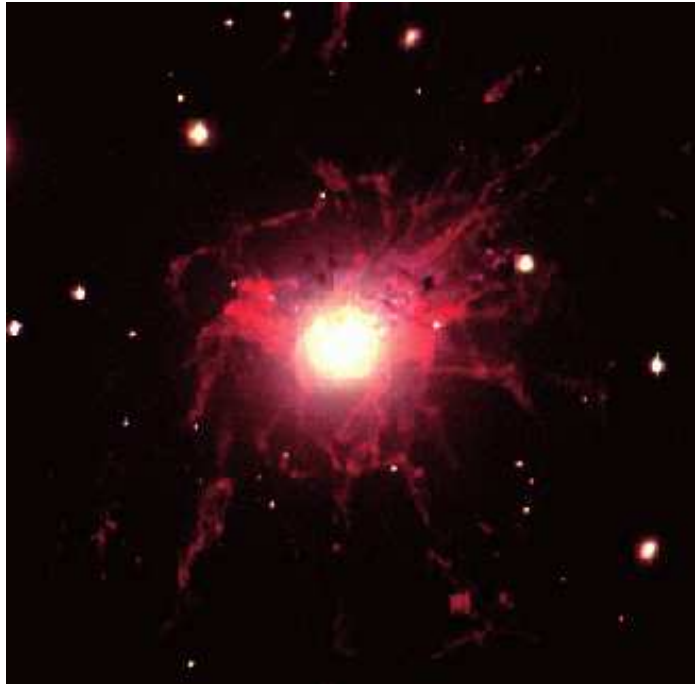}{6.0in}{0}{80}{80}{-250}{-100}
\caption{Combined color image of NGC~1275 using the B,
R, and H$\alpha$ WIYN filters.  The ionization structure
can clearly be seen, including the tangential filaments oriented in
the east-west direction and the longer 50\hub kpc filaments seen projecting 
towards the north and south. {\bf Note
that this figure is best seen in color.}}
\end{figure}

\clearpage

\begin{figure}
\plotfiddle{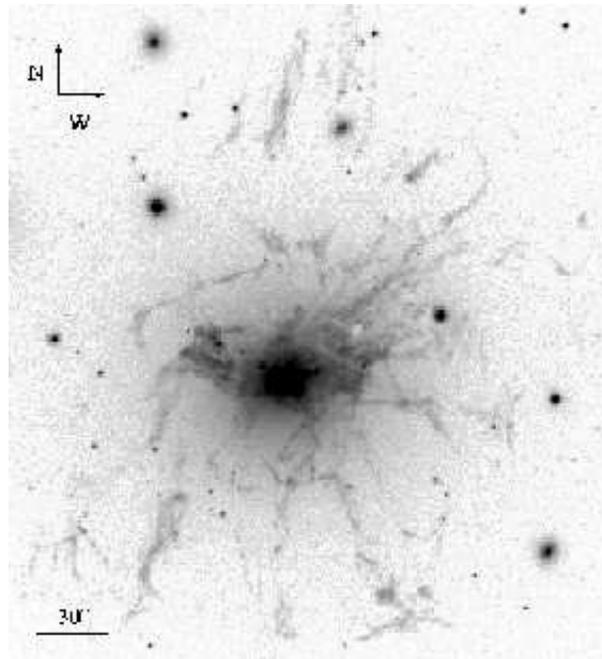}{6.0in}{0}{80}{80}{-250}{-100}
\vskip 0.5in
\caption{An H$\alpha$ image, before continuum subtraction, showing the 
low-velocity system's ionized gas filaments.}
\end{figure}

\clearpage

\begin{figure}

\plotfiddle{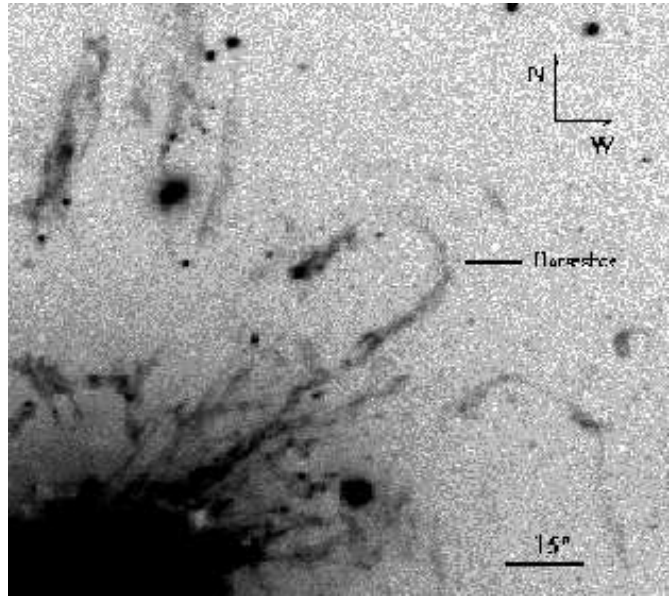}{6.0in}{0}{80}{80}{-250}{-100}
\caption{Close up of Figure 2 showing an example of
a radial filament.  This particular filament reverses direction, a possible
example of a magnetic Rayleigh-Taylor instability.}
\end{figure}

\clearpage

\begin{figure}
\vskip -2in
\plotfiddle{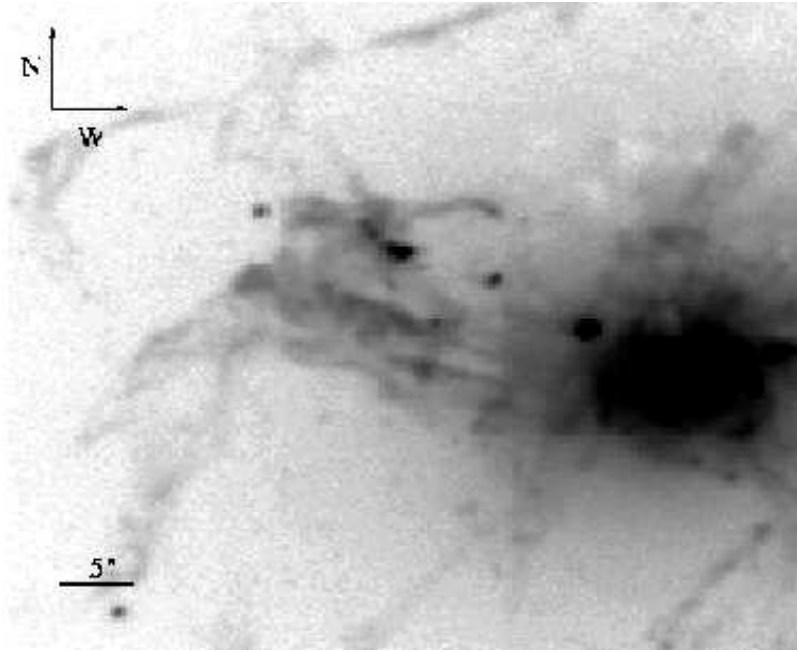}{7.0in}{0}{80}{80}{-250}{-100}
%\vskip 1.0in
\caption{Close up of Figure 2 showing examples of
tangential filaments that are possibly the instability interface between
the radio plasma and the cooling flow X-ray gas.}
\end{figure}

\clearpage

\begin{figure}
\vskip -1in
\plotfiddle{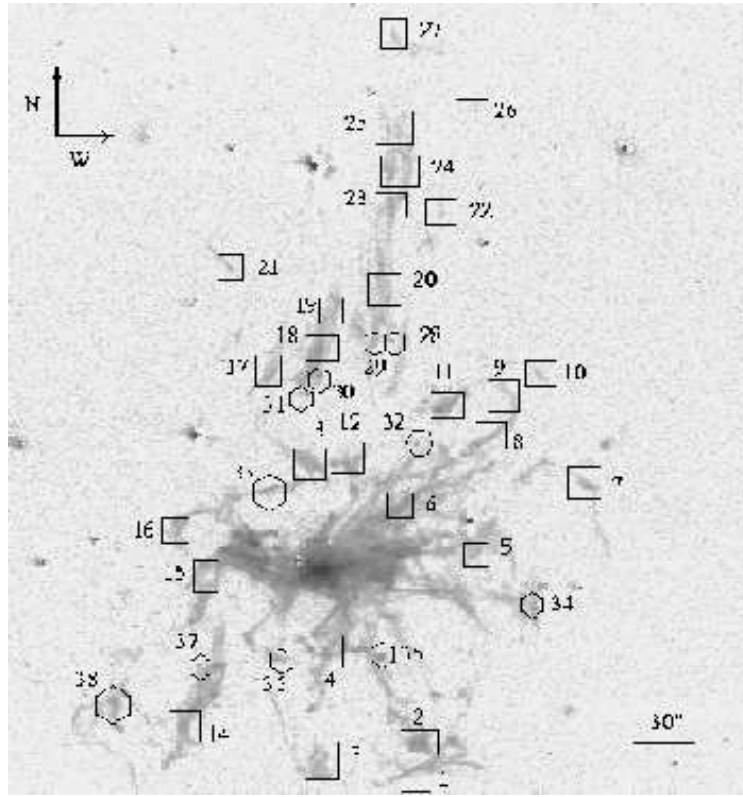}{7.0in}{0}{80}{80}{-250}{-100}
\vskip 0.3in
\caption{Continuum subtracted H$\alpha$ map of NGC~1275 showing the 
locations of areas used to measure the intensities of emission features listed
in Table 1.  The hexagons are for H$\alpha$ knots, and boxes are for diffuse
areas. {\bf Note: to properly view this image, please obtain the version
of this paper at: http://www.astro.wisc.edu/$\sim$chris/pera.pa. }}
\end{figure}

\clearpage

\begin{figure}
\vskip -1.5in
\plotfiddle{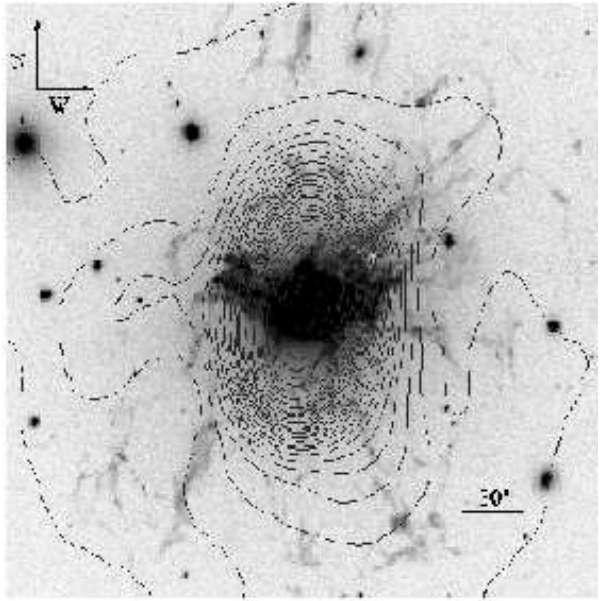}{7.0in}{0}{80}{80}{-250}{-100}
\vskip -0.4in
\caption{H$\alpha$ emission with 1320 MHz radio contours superimposed.}
\end{figure}

\clearpage

\begin{figure}
\plotfiddle{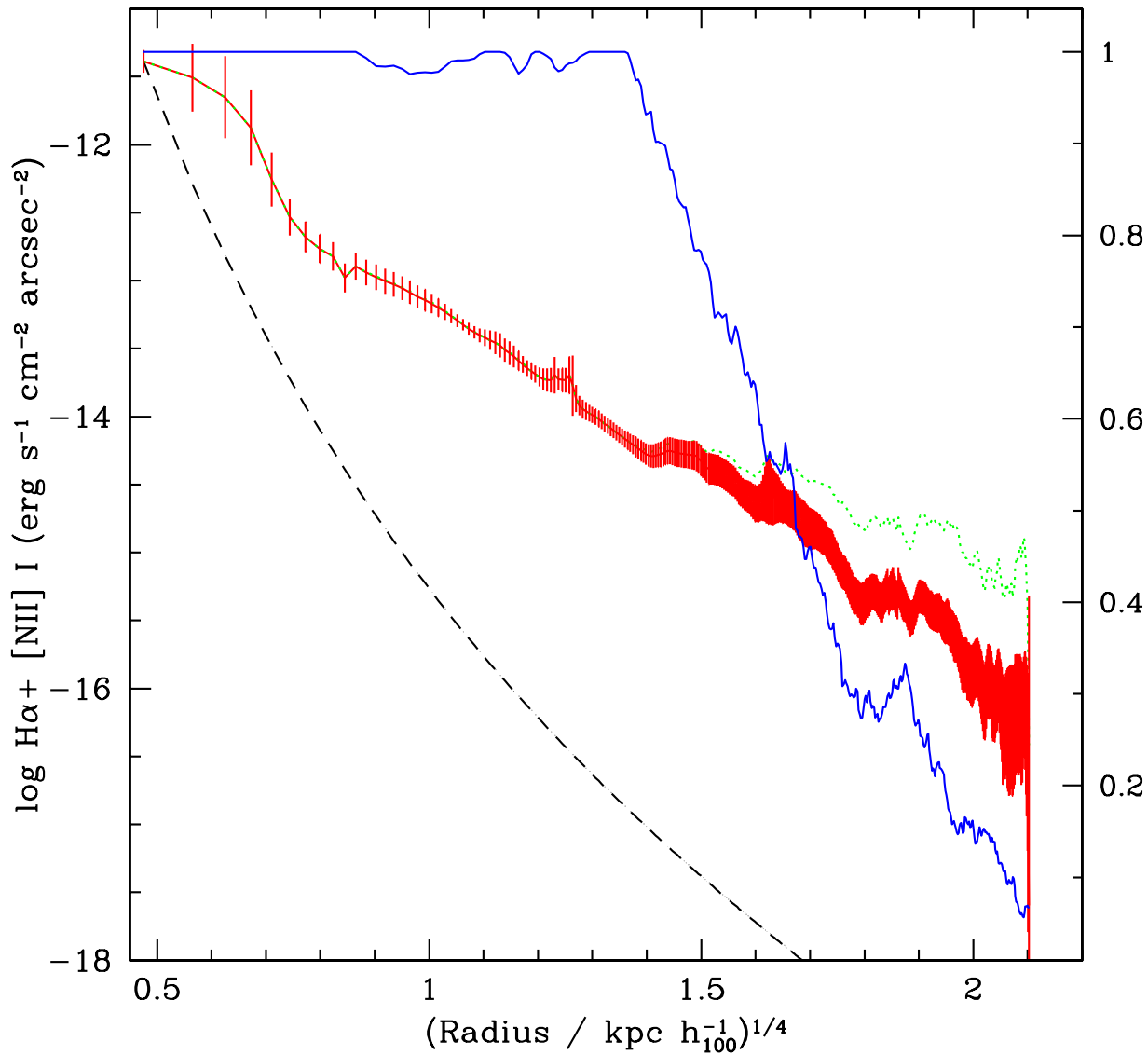}{7.0in}{0}{80}{80}{-250}{-100}
\vskip -0.3in
\caption{The observed mean H$\alpha$ intensity profile of the ionized
gas around NGC~1275, as measured in elliptical apertures, is shown as a red 
line with error bars, plotted on the left intensity scale.  The dashed line
shows the predicted intensity with projected radius, R,
for a central ionization
source model with the only losses being r$^{-2}$ dimming. Using the scale
on the right, we show the upper limit to the projected ionized gas filling 
factor
versus radius (blue solid line). The inner 3.8\hub ~kpc is almost
completely filled with overlapping filaments, while the outer parts of the
system are very sparsely occupied. The green dotted line above the mean
intensity profile at larger radii illustrates the brightness that would be
observed at each radius if the ionized gas projected filling factor were
unity.  {\bf Note that this figure is best seen in color.}}
\end{figure}

\clearpage

\begin{figure}
\plotfiddle{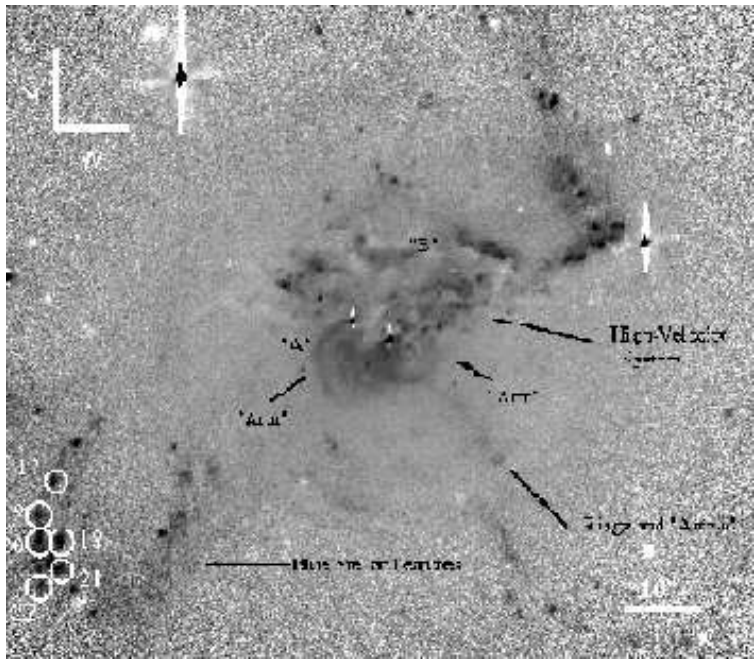}{7.0in}{0}{80}{80}{-250}{-100}

\caption{(B$-$R) color image of the central parts of NGC~1275.
Darker is bluer.  Several features stand out in the image, including the
HV system and its associated dust.  A blue streak of star clusters can be
seen towards the north-west and south-east of the center.   South of the center
of NGC~1275 are four rings, tidal arms, and an arrow feature, all
indications of recent merger activity (see text). The labeled clusters have
photometric properties listed in Table 2. {\bf Note: to properly view this 
image, please obtain the version  of this paper
at: http://www.astro.wisc.edu/$\sim$chris/pera.pa. }}
\end{figure}

\clearpage

\begin{figure}
\vskip -0.5in
\plotfiddle{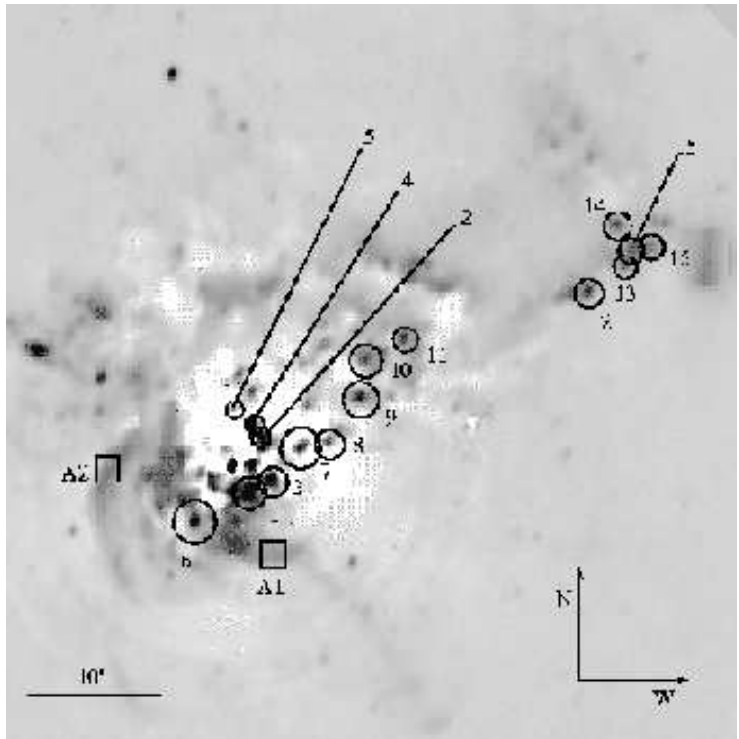}{7.0in}{0}{80}{80}{-250}{-100}
\caption{Inner stellar features of the NGC~1275 system.  Star
clusters whose photometry is listed in Table 2 are circled.  The
areas where features of the `arms' are measured are labeled as boxes and
A1, A2.  This figure can be compared with Figure 8 by the location of
the two `arms'. {\bf Note: to properly view this image, please obtain the
version of this paper at: http://www.astro.wisc.edu/$\sim$chris/pera.pa. }}
\end{figure}

\clearpage

\begin{figure}
\plotfiddle{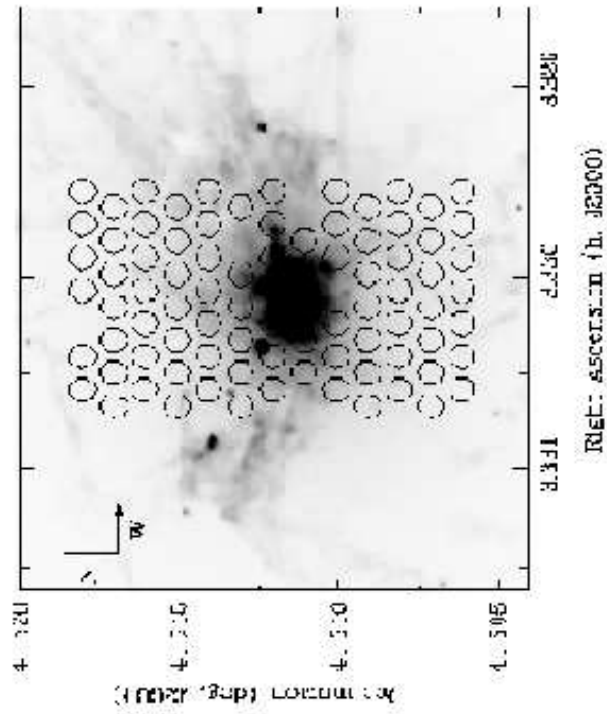}{7.0in}{0}{80}{80}{-250}{-100}
\caption{Fiber positions of Densepak across the
center of NGC~1275.}
\end{figure}

\clearpage

\begin{figure}
\vskip -1in
\plotfiddle{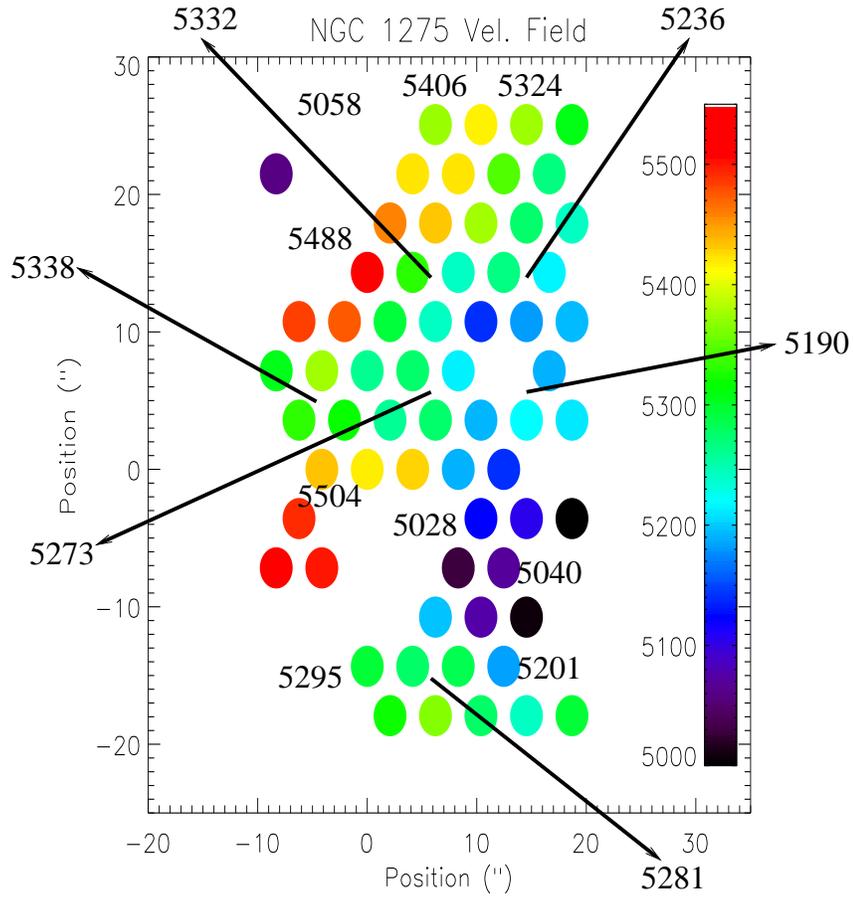}{7.0in}{0}{80}{80}{-250}{-100}
\vskip -0.3in
\caption{Velocity field of the central 45\arcsec\, of
NGC~1275.  The velocity range is about 500 \kms, with an apparent
rotation from south-east to north-west at a position angle of
120\deg.  The gradient is clear when imaged is viewed in
color, although the gradient can be seen by examining the average
velocities for 10\arcsec\, square boxes which are labeled. {\bf Note
that this figure is best seen in color.}}
\end{figure}

\clearpage

\begin{figure}
\vskip -2in
\plotfiddle{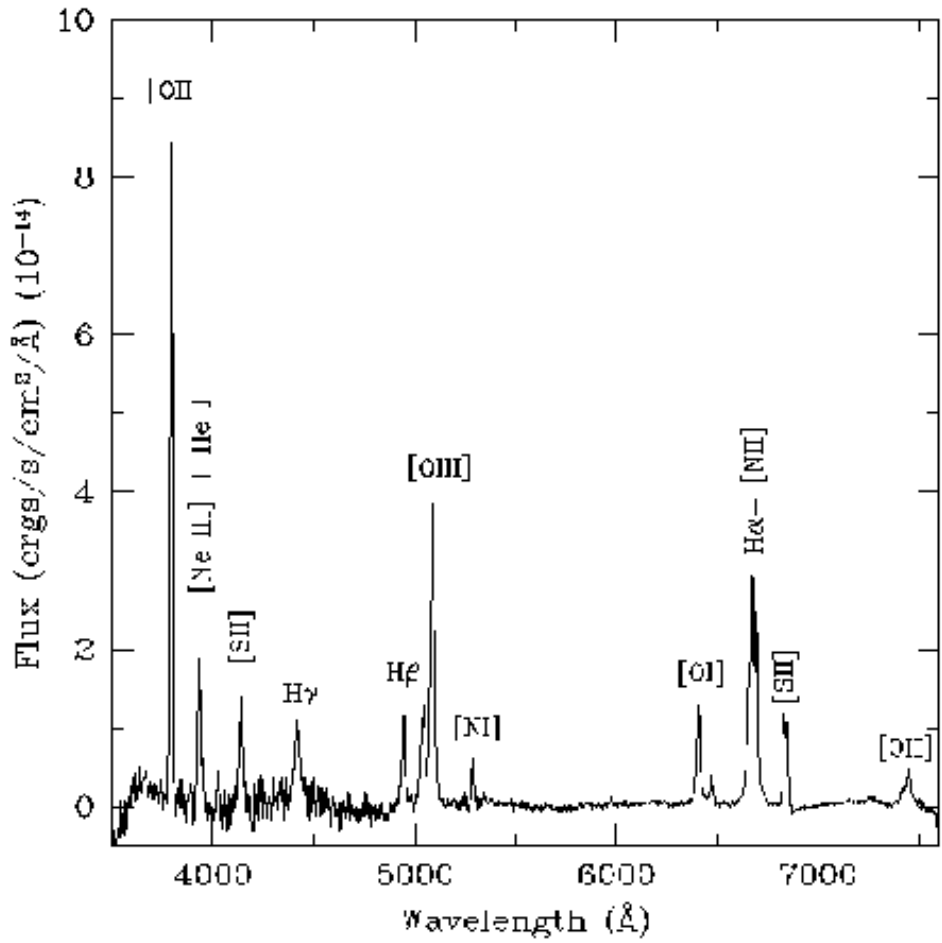}{7.0in}{0}{80}{80}{-250}{-100}
\caption{Long-slit spectrum of an H$\alpha$ filament near the
center of NGC~1275.  Note the combination of neutral and ionized
species, as discussed in the text.}
\end{figure}

\begin{figure}
\plotfiddle{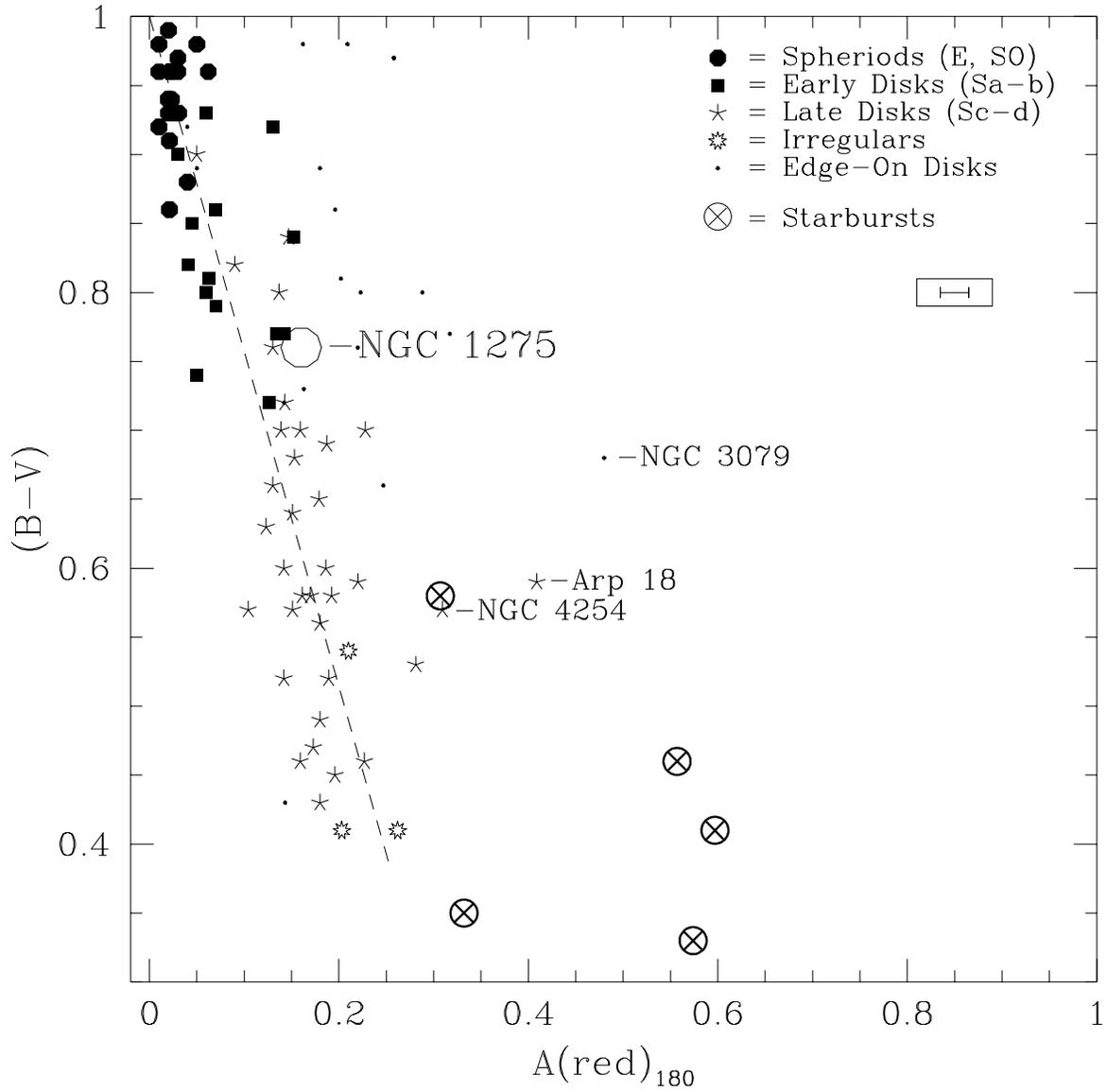}{7.0in}{0}{80}{80}{-250}{-100}
\caption{Color-Asymmetry Diagram (CAD) for a sample of
galaxies, including the position of NGC~1275.  This diagram and the ripple
features in Figures 8 and 9 reveals that some type of
merger took place, probably recently.}
\end{figure}